\documentclass[11pt,letterpaper,twocolumn]{article}
\usepackage{times}
\usepackage[top=2.25cm,bottom=2.4cm,left=2cm,right=2cm]{geometry}

\usepackage[colorlinks,urlcolor=blue]{hyperref}
\usepackage{graphicx}
\usepackage{mathtools}
\usepackage{dsfont}
\usepackage{textgreek}
\usepackage[normalem]{ulem}
\usepackage[font=small]{caption}
\usepackage[square,numbers,sort&compress]{natbib}
\usepackage{balance}
\usepackage{dblfloatfix}

\makeatletter
\DeclareRobustCommand*\textsubscript[1]{\@textsubscript{\selectfont#1}}
\def\@textsubscript#1{{\m@th\ensuremath{_{\mbox{\fontsize\sf@size\z@#1}}}}}
\makeatother

\usepackage[dvipsnames]{xcolor}
\newenvironment{sciabstract}{
\begin{quote} \bf}
{\end{quote}}

\title{Quantum Mechanics of Proteins in Explicit Water: The Role of Plasmon-Like Solute-Solvent Interactions}

\author{{Martin St\"{o}hr} and {Alexandre Tkatchenko}$^\ast$\\
\\
\normalsize{Physics and Materials Science Research Unit, University of Luxembourg,}\\
\normalsize{L-1511 Luxembourg, Luxembourg}\\
\\
\normalsize{$^\ast$To whom correspondence should be addressed; E-mail: \href{mailto:alexandre.tkatchenko@uni.lu}{alexandre.tkatchenko{@}uni.lu}}}

\date{}

\usepackage{titling}
\begin{document}

\baselineskip20pt

%\maketitle

\twocolumn[
  \begin{@twocolumnfalse}
    \maketitle
    \vspace{-1.25cm}
    \baselineskip16pt
    \begin{sciabstract}
Quantum-mechanical van der Waals dispersion interactions play an essential role for both \textit{intra}-protein and protein-water interactions -- the two main driving forces for the structure and dynamics of proteins in aqueous solution.
Typically, these interactions are only treated phenomenologically via pairwise potential terms in classical force fields.
Here, we use an explicit quantum-mechanical approach based on density-functional tight-binding with the many-body dispersion formalism, which allows us to demonstrate the unexpected relevance of the many-body character of dispersion interactions for protein energetics and the protein-water interaction.
In contrast to commonly employed pairwise approaches, many-body effects significantly decrease the relative stability of the native state in the absence of water.
In an aqueous environment, the collective character of the protein-water van der Waals interaction counteracts this effect and stabilizes native conformations and transition states.
This stabilization arises due to a high degree of delocalization and collectivity of protein-water dispersion interactions, suggesting a remarkable persistence of long-range electron correlation through aqueous environments.
Our findings are exemplified on prototypical showcases of proteins forming \textbeta-sheets, hairpins, and helices, emphasizing the crucial role of plasmon-like solute-solvent interactions in biomolecular systems. 
    \vspace{0.55cm}
    \end{sciabstract}
  \end{@twocolumnfalse}
]
\baselineskip15pt

\section*{Introduction}
Water is an essential basis of life. It provides the environment in which the biomolecular machinery can exist and function.
By screening and stabilizing static electronic multipoles, water significantly alters the structure, stability, and dynamics of biomolecules~\cite{Xu1999,Patriksson2007,Bellissent-Funel2016}.
The favorable exposure of moieties with static electronic multipoles to water and the corresponding burying of non-polar residues into a hydrophobic core, is also an important, two-fold driving force for protein folding:
First, the (short-range) interaction with the aqueous environment is optimized and, second, the disruption of the dynamic hydrogen bond network of the surrounding water by hydrophobic residues is minimized~\cite{Kauzmann1959,Chandler2005,Bellissent-Funel2016}.
While the importance of this ``hydrophobic effect" and the pivotal role of water for biomolecular systems is under no dispute~\cite{Kauzmann1959,Honig1995,Chandler2005,Fersht2017}, the underlying fundamental physics of solvated (bio)molecular systems is still not fully explored and understood.
In particular, here we focus on the quantum-mechanical nature of solute-solvent interactions.
It has already been shown that polarization effects and the many-body character of bonded interactions and hydrogen bond networks play an important role for solvated systems~\cite{Dannenberg2005,Shvab2013,Rossi2015,Bellissent-Funel2016,Cisneros2016}, but also long-range van der Waals (vdW) dispersion interactions form an essential component for water and for both \textit{intra}-protein and protein-water interactions.
This vdW component, however, has not been investigated in full detail nor on a fundamental level up to now.
In the present study, we address exactly this intricate issue and find that these quantum-mechanical interactions can account for up to 30\,\% of the total solvation energy.
Together with their essential role for \textit{intra}- and inter-protein interactions, this calls for a more complete microscopic understanding of vdW dispersion forces under physiological conditions, which is imperative to shed light on the physics of proteins in aqueous solvation.

Non-covalent vdW dispersion interactions arise from instantaneous, correlated fluctuations of the electron density and, hence, are inherently quantum-mechanical and many-body in nature.
However, current solvation models or molecular mechanics force fields include them only in a phenomenological manner via pairwise-additive potentials.
Recent studies demonstrate that, for a variety of systems, including polypeptides in gas phase~\cite{Rossi2015,Schubert2015a}, such approximations can fundamentally fail and that one has to explicitly account for the intrinsic many-body character of vdW interactions~\cite{Tkatchenko2012,Gobre2013,Ambrosetti2014,Reilly2014,Reilly2015,Schubert2015a,Hermann2017,Hoja2017,Dobson2014,Kronik2014} -- even for the properties of pristine water~\cite{Jones2013a} and its surfaces~\cite{Sokhan2015}.
Moreover, pairwise or phenomenological models only provide a very approximate description of the energetic aspects of vdW forces, but no description of the underlying quantum mechanics.
Such insights, however, can be vital to comprehend and conceptually understand vdW interactions as recently illustrated for hybrid and nanostructured systems~\cite{Gao2016,Ambrosetti2016} or \textpi--\textpi\,stacked molecules~\cite{Hermann2017}.
It is important to note that the dielectric permittivity of water has a value around 2.3~\cite{Yada2009} at the frequencies of the electronic fluctuations, which are responsible for dispersion interactions, \textit{i.e.} at several hundred terahertz (THz).
Therefore, in contrast to static electronic multipoles, vdW interactions are not strongly screened by aqueous environments and, thus, can give rise to long-range interactions also in solvated systems.
As such, long-range correlation forces may play an important role for the long-range ordering often observed in biological systems or form the quantum-mechanical basis for the emergence of coherent molecular vibrations~\cite{Reilly2014,Folpini2017}.
Such collective nuclear behavior has been proposed to play an important role in long-distance recognition among biological macromolecules~\cite{Froehlich1968,Acbas2014,Preto2015,Nardecchia2018}.
Within the conventional view of solvated proteins, however, the basis for long-range recognition under physiological conditions is still controversially discussed.
Recent studies also suggest connections between collective electronic fluctuations -- the basis of vdW dispersion interactions -- and enzymatic action on DNA~\cite{Kurian2016,Kurian2018} or pharmaceutical activity~\cite{Craddock2017}.

In this work, we aim at a more complete microscopic description of solvated proteins and report an explicitly quantum-mechanical study, which accounts for many-body dispersion interactions as well as for electrostatics, polarization, and hydrogen bonding.
Molecular mechanics approaches are often found to reliably capture the latter three, (semi-)classical interactions.
This, however, usually holds true only for certain conditions (typically designed for the liquid phase at room temperature and ambient pressure), which shows that the physical description is incomplete.
As especially the quantum-mechanical, non-local dispersion interactions are often described insufficiently and inconclusively by conventional approaches~\cite{Baldauf2015,Schubert2015a}, we focus on a comprehensive description of vdW interactions and collective electronic behavior to highlight the role of water for the vdW energetics of prototypical fast-folding proteins.
In this context, previous studies have pointed out that traditional molecular mechanics potentials and water models likely provide an unbalanced description of vdW interactions for proteins in water, which typically results in an over-compaction of unfolded states~\cite{Knott2012,Best2014,Piana2015}.
Typically, this unbalanced description is approached by adapting the pairwise vdW interaction coefficients for the \textit{intra}-protein, water-water, or protein-water interaction.
In this work, we seek to understand the potential fundamental basis for the failure of the traditional models in a bottom-up approach.
The calculations have been carried out using a combined approach~\cite{Stoehr2016} of the Density-Functional Tight-Binding (DFTB) method~\cite{Porezag1995,Elstner1998} and accurate \textit{ab initio} dispersion models, which allows for a robust, albeit approximate, quantum-mechanical treatment on an atomistic level.
In fact, an \textit{ab initio} description of vdW interactions is the only way to study the role of the solvent, as force field methods are typically strongly limited in their transferability between gas and liquid phase due to their high degree of parameterization.
In our study, we focus on a comparison of a pairwise-additive description of vdW interactions and an accurate, quantum-mechanical many-body treatment.
The pairwise vdW models are represented by the vdW(TS)~\cite{Tkatchenko2009} as well as Grimme's D2~\cite{Grimme2006} and D3~\cite{Grimme2010,Grimme2011} approaches.
Such an approximate, pairwise formalism represents the basis for the standard phenomenological description of long-range correlation in biomolecular simulations via Lennard-Jones potentials.
For comparison, we study the vdW interaction within the Many-Body Dispersion (MBD) formalism~\cite{Tkatchenko2012,Tkatchenko2013}, which accounts for the many-body character of vdW dispersion interactions to infinite order in perturbation theory within an interatomic framework and has been proven to provide quantitative improvements and a better qualitative understanding compared to the pairwise-additive approximation in numerous studies~\cite{Tkatchenko2012,Gobre2013,Ambrosetti2014,Reilly2014,Reilly2015,Schubert2015a,Hermann2017,Hoja2017}.
Yet, its computational efficiency together with modern implementations and the ever-growing availability of computational resources, allows for treatment of systems consisting of several thousands of atoms as in the case of proteins in explicit solvent.
The MBD formalism also represents a model for collective electronic fluctuations~\cite{Gao2016,Ambrosetti2016,Hermann2017} -- a molecular analogue to the plasmon pseudo-particle in metallic systems -- which we will use to further characterize the protein-water interaction.

%RESULTS
\begin{figure*}[hbt!]
\centering
\includegraphics[scale=0.8]{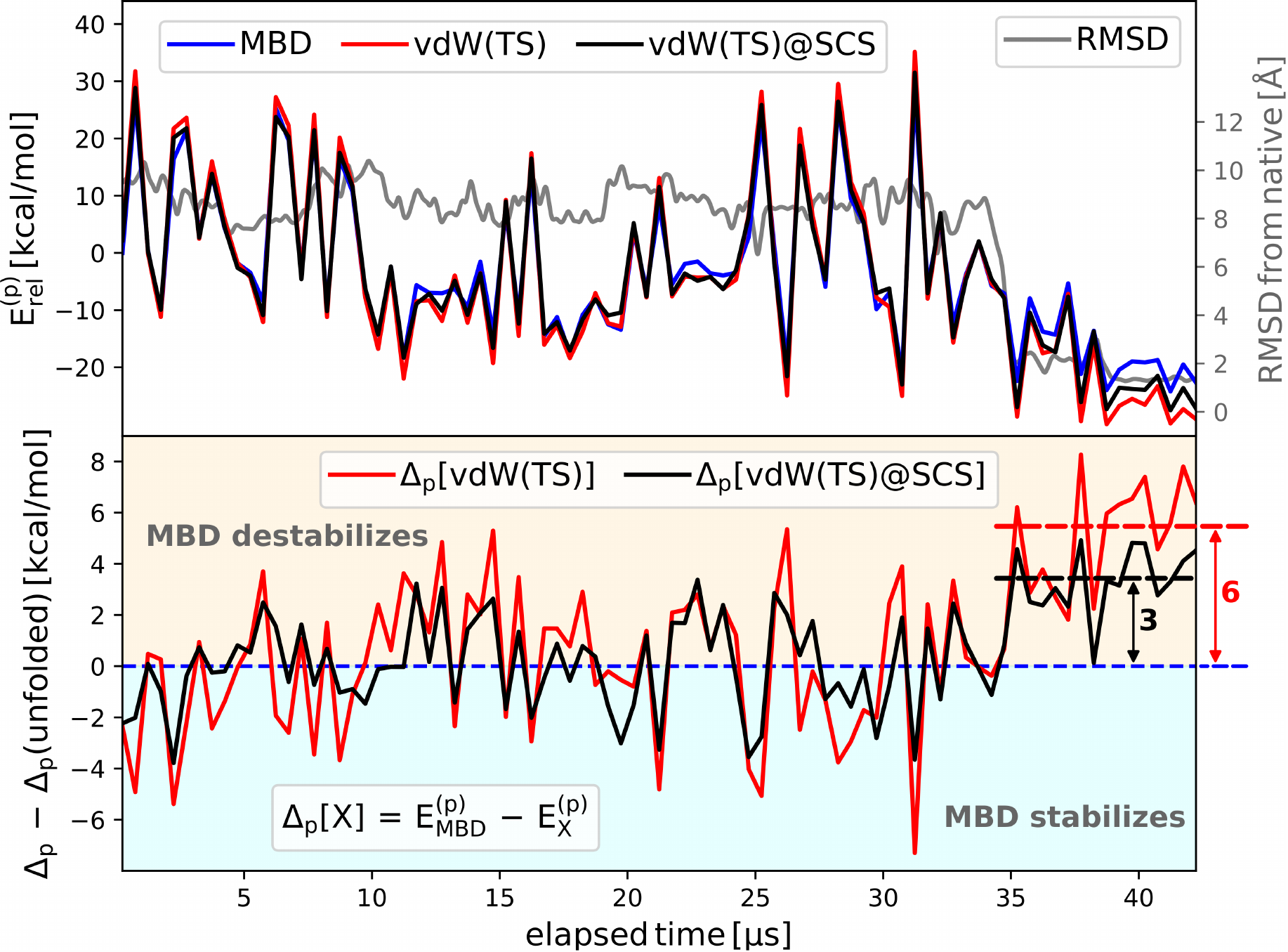}
\caption{top: van der Waals energy of Fip35 Hpin1 WW-domain in solvated geometry without solvent. bottom: Beyond-pairwise contributions, as given by difference between the many-body formalism MBD and the pairwise treatments vdW(TS) and vdW(TS)@SCS, \textit{i.e.} vdW(TS) with self-consistent screening. For a more comprehensive version, see Supplementary Materials.}
\label{Fig:Ep_vdW-FiP35WW}
\end{figure*}
\section*{Results}
We exemplify our findings in detail for the Fip35 Hpin1 WW domain~(Fip35-WW)~\cite{Liu2008} and further showcase their general validity at hand for the \textit{de novo} Chignolin variant ``cln025"~\cite{Honda2008} and the fast-folding Nle/Nle double mutant of the villin headpiece (HP35-NleNle)~\cite{Kubelka2006}.
The folding trajectories of the three proteins have been obtained in atomistic detail and explicit solvent in previous molecular dynamics simulations by Shaw~\textit{et~al.}~\cite{Shaw2010}, Lindorff-Larsen~\textit{et~al.}~\cite{Lindorff-Larsen2011}, and Ensign~\textit{et~al.}~\cite{Ensign2007,Ensign2007a}, respectively.

%gas phase
\subsection*{\textit{Intra}-protein interactions and many-body dispersion effects}
We start out by investigating the Fip35-WW trajectory by artificially removing the surrounding solvent from a molecular dynamics trajectory in explicit water, \textit{i.e.} we first focus on \textit{intra}-protein interactions, where dispersion forces represent one of the main sources of interaction within the protein core.
Accordingly, we observe an increased magnitude of the vdW energy while this core is being formed and particularly during the hydrophobic collapse in all applied dispersion models (see Fig.~\ref{Fig:Ep_vdW-FiP35WW} and Supplementary Material).
Notably, in comparison to the results obtained within the pairwise approaches, many-body dispersion effects consistently decrease the relative stability of the native state for the isolated protein by 6~kcal/mol on average, \textit{cf}. Fig.~\ref{Fig:Ep_vdW-FiP35WW}(bottom).
The outliers of this general behavior observed around 15 and 26~{\textmu}s correspond to transient, partly folded intermediates.
The relative destabilization by beyond-pairwise contributions can be explained by an overestimation of the \textit{intra}-core vdW interactions (``overcorrelation") in the pairwise approximation.
By reducing the interaction to pairwise-additive potentials, a two-body formulation assumes ideal correlation between all pairs of atoms and with that, neglects the complex geometrical arrangement within the protein core. 
Such geometrical constraints limit the emergence of correlated fluctuating dipole patterns and thus lower the interaction energy as already observed for a wide variety of systems~\cite{Gobre2013,Ambrosetti2014,Reilly2015,Hermann2017,Yang2017}.
For small peptides such effects have been found to be mostly negligible~\cite{Baldauf2015,Rossi2014a}.
Our findings show that for larger biomolecules, however, a many-body treatment of vdW interactions is indeed essential, which is in line with the findings of Schubert \textit{et~al.} for 20-residue peptides~\cite{Schubert2015a}.

In the MBD formalism, we make use of a two-step procedure: We obtain effective, screened atomic polarizabilities from self-consistent electrodynamic screening to account for the presence of multiple fluctuating dipoles in the system and then solve a many-body Hamiltonian, which is defined in terms of these polarizabilities, to capture many-body vdW interactions.
To study the effect of each step, we combined vdW(TS) with the self-consistent screening procedure.
In this variant, which we refer to as vdW(TS)@SCS, screened interaction coefficients enter the pairwise-additive potentials instead of the hybridized chemical analogue used in vdW(TS).
In this way, we account for the effects on atomic polarizabilities due to the local field of the surrounding dipoles, but do not include long-range many-body interactions.
Within vdW(TS)@SCS we already capture some part of the destabilization of native states amounting to 3~kcal/mol (\textit{cf}. Fig.~\ref{Fig:Ep_vdW-FiP35WW}).
Thus, half of the overstabilization in vdW(TS) is from neglecting the presence of multiple dipoles in the system and half from the many-body character of dispersion interactions.
This also implies that for a proper description of gas phase proteins, one has to account for the screening of polarizabilities and the many-body nature of vdW interactions.

%vdW solvation energy
\subsection*{van der Waals solvation energy}
\begin{figure*}[htb!]
\centerline{\includegraphics[scale=0.8]{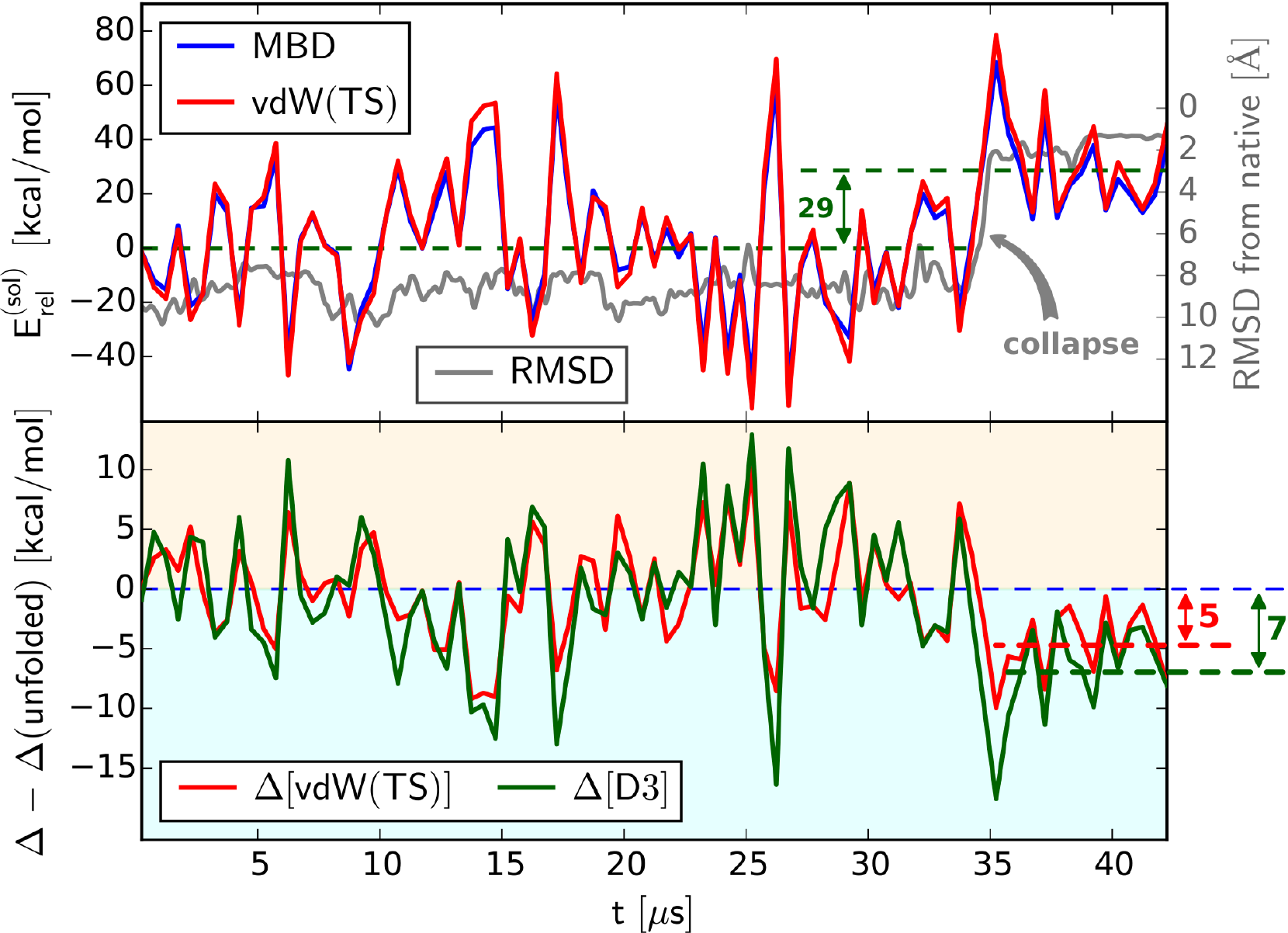}}
\caption{Relative vdW solvation energy, $E_{\rm rel}^{\rm (sol)}$, during the folding process of the Fip35 Hpin1 WW domain. \textbf{top}: backbone root-mean-square deviation from final conformation illustrating the hydrophobic collapse around 35~{\textmu}s (gray). The vdW contribution to the relative solvation energy is shown for the pairwise vdW(TS) model (red) and MBD (blue). \textbf{bottom}: Difference in the relative stabilization by the solvent between MBD and the pairwise vdW(TS) and D3 referenced to the unfolded state.}
\label{Fig:Esol_vdW-FiP35WW}
\end{figure*}

Fig.~\ref{Fig:Esol_vdW-FiP35WW}(top) shows the vdW contribution to the solvation energy obtained with MBD and the vdW(TS) model, as defined by
\begin{equation}
E(\mathrm{sol}) = E_{\mathrm{vdW}}[\mathrm{ps}] - E_{\mathrm{vdW}}[\mathrm{p}] - E_{\mathrm{vdW}}[\mathrm{s}]\quad,
\label{eq:Esol}
\end{equation}
with $\mathrm{ps}$ referring to Fip35-WW in solvation, $\mathrm{p}$ to Fip35-WW in gas phase, and $\mathrm{s}$ to the pristine solvent.
As an artifact of the above-mentioned over-correlation within the pairwise approach, we find a consistent overestimation of the dispersion contribution in vdW(TS).
In terms of the relative solvation energy, however, pairwise and many-body treatment show the same general trend, which qualitatively follows the inverse root-mean square deviation (RMSD) from the native state with a step coinciding with the collapse of the protein into the native, more globular shape.
This finding can be explained by the removal of hydrophobic residues from the protein-water interface and thus decreasing their interaction with the solvent.
The average dispersion contribution to the solvation energy drops by 29~kcal/mol (15\,\%) at the hydrophobic collapse.
The step-like behavior of the vdW solvation energy along the trajectory is even more distinct than for the \textit{intra}-protein vdW interaction in the gas phase and almost resembles a two-state model of folded and unfolded states.
As such, the vdW solvation energy prominently captures the protein's collapse and, thus, represents a valid descriptor for the folding process.
This feature has been found for all dispersion models considered here: the MBD formalism and the pairwise approaches vdW(TS), D2, and D3.

Comparing many-body and pairwise treatment of dispersion interactions, Fip35-WW does no longer feature a consistent change in the relative stability of native versus non-native conformations once embedded in an aqueous environment.
Thus, beyond-pairwise effects in the protein-water vdW interaction stabilize folded conformations.
Correspondingly, we see a clear increase in the relative vdW solvation energy for native and native-like states, when comparing the pairwise models to MBD (5~kcal/mol for vdW(TS), 7~kcal/mol for D3).
This shift is due to the lack of a systematic many-body (de-)stabilization in the total vdW energy of solvated Fip35-WW and the water box during the whole folding trajectory, combined with an inversion of the behavior observed for the isolated protein shown in Fig.~\ref{Fig:Ep_vdW-FiP35WW} (see Eq.~\ref{eq:Esol} and Supplementary Materials).
This implies that the protein-water interaction compensates for the destabilization of native states via many-body dispersion effects, observed \textit{in vacuo}.
In summary, besides screening permanent electronic multipoles, water also provides the necessary environment to stabilize native conformations via beyond-pairwise vdW interactions, which counteracts the destabilizing effect that such many-body terms have on the \textit{intra}-protein interaction.

%plasmon-like interactions
\subsection*{The plasmon-like character of protein-water vdW interactions}
\begin{figure*}[!b]
\centerline{\includegraphics[scale=.99]{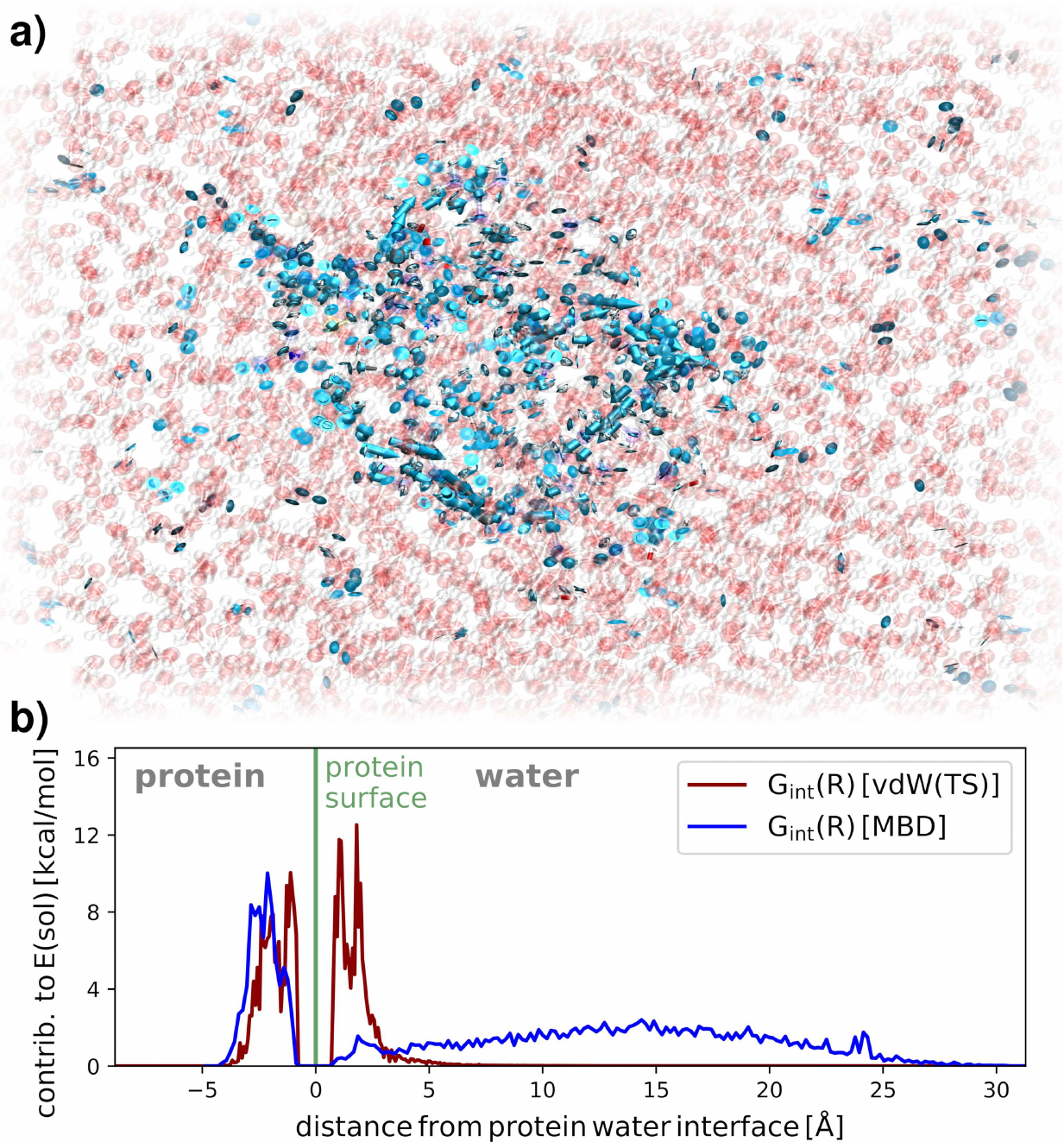}}
\caption{\textbf{a)} Illustration of low-frequency plasmon-like fluctuations in solvated Fip35 Hpin1 WW domain, which show the largest contribution to the protein-water interaction. The arrows depict the direction of simultaneous electron density deformations (eigenmode of the electron density). \textbf{b)} Contributions to the vdW solvation energy within the pairwise vdW(TS) approach and the many-body dispersion formalism (MBD) as radial distribution functions.}
\label{Fig:displacements-FiP35WW}
\end{figure*}
As has been shown previously, MBD also provides a model for the intrinsic electronic fluctuations~\cite{Gao2016,Ambrosetti2016,Hermann2017}: 
In analogy to nuclear quantum effects, the electronic fluctuations, which ultimately give rise to vdW interactions, can be understood as the zero-point oscillations around the average electron density.
The MBD formalism gives access to an orthonormal decomposition of this zero-point fluctuation, which can be interpreted as ``eigenmodes'' of the electron density.
A detailed analysis of these electronic eigenmodes reveals, that the number of very localized, high-frequency oscillations, formerly mainly located on the solute, significantly decreases upon coupling to the surrounding water.
This implies a delocalization of electronic fluctuations and an increase of the collectivity of electronic behavior.
This plasmon-like character and the delocalization over protein and solvent form the fundamental reason for the stabilization of native states via many-body dispersion effects in the protein-water interaction.
The role of the surrounding solvent can be seen as providing weakly structured polarizable matter, which attenuates the destabilizing many-body effects observed \textit{in vacuo} for native and partially folded states.
To gain further insight into the characteristics of protein-water vdW interactions, we additionally obtain the contribution of individual electronic fluctuations to the solvation energy:
The main contributions to the vdW solvation energy are due to highly collective, plasmon-like electronic oscillations around 450 THz ($\hat{=}$ 670\,nm), such as the one depicted in Fig.~\ref{Fig:displacements-FiP35WW}a).
Their contribution to the total protein-water interaction is 14--16\,\% ($\approx$ 41~kcal/mol) and exceeds 20\,\% ($\approx$ 6~kcal/mol) for relative solvation energies.\\

\begin{figure*}[!b]
\centerline{\includegraphics[scale=1]{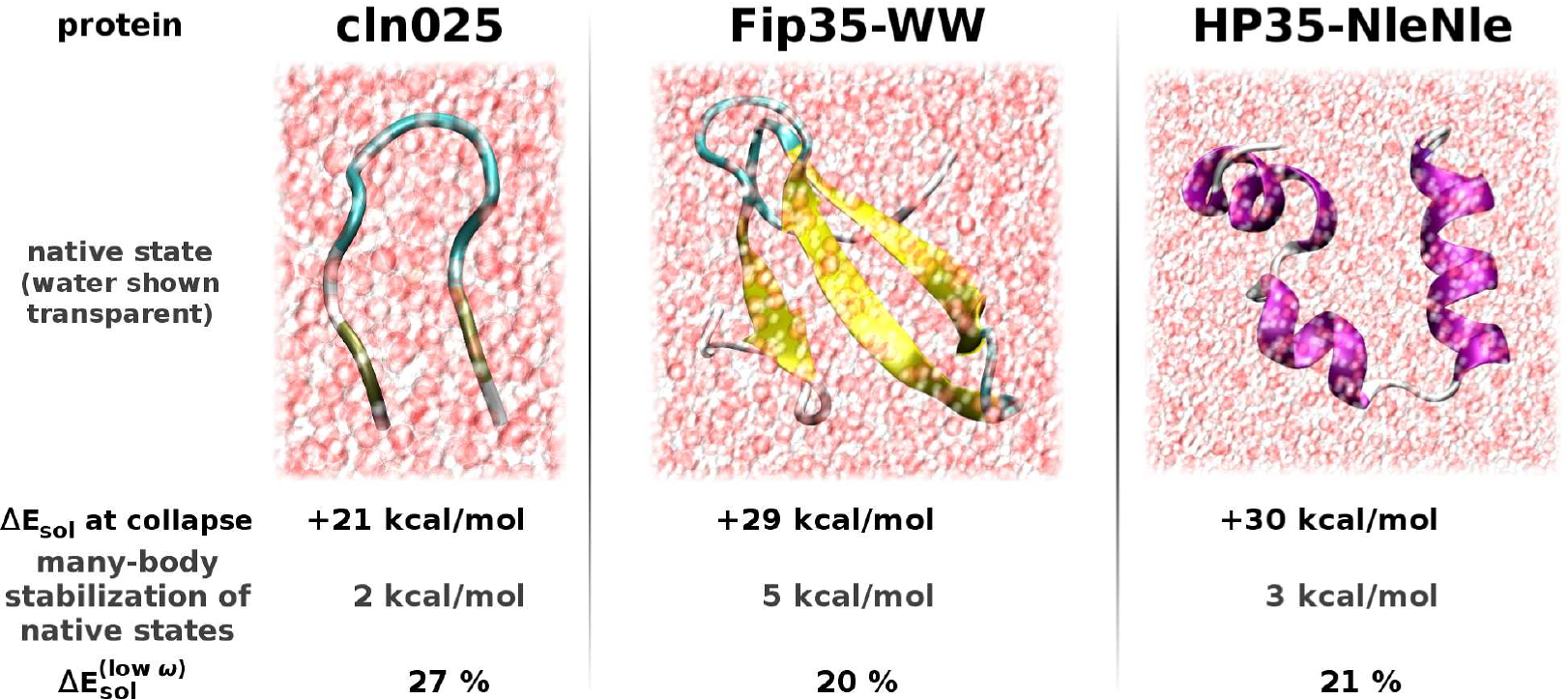}}
\caption{Characteristics of protein-water dispersion interactions: Independent of the secondary structure, the van der Waals solvation energy captures the hydrophobic collapse in form of a 20--30~kcal/mol jump (``$\Delta E\rm{_{sol}}$ at~collapse") and many-body protein-water van der Waals interactions consistently stabilize native states in solvation. Low frequency, collective electronic fluctuations contribute significantly to the relative solvation energy ($\Delta E\rm{_{sol}^{(low\,\omega)}}$) in all cases.}\label{Fig:Summary}
\end{figure*}
As demonstrated in Fig.~\ref{Fig:displacements-FiP35WW}a), such fluctuations commonly feature large charge displacements along the polarizable protein backbone coupled to electronic fluctuations throughout the solvent.
Notably, these plasmon-like fluctuations reach from the protein backbone inside the hydrophobic core far into the aqueous environment.
Comparing the radial distribution of the contributions to the vdW solvation energy between the pairwise vdW(TS) and MBD models, as shown in Fig.~\ref{Fig:displacements-FiP35WW}b), reveals a striking difference in the interaction range within the two treatments:
In the pairwise model, the contribution of solvent atoms to the vdW solvation energy subsides beyond 6\,\AA, \textit{i.e.} roughly twice the sum of the vdW radii of carbon and oxygen.
Accounting for many-body dispersion, on the other hand, shows that electronic correlation between the protein and solvent atoms up to 25\,\AA\:from the protein-water interface is still relevant for the protein-water interaction.
This reflects the weakness of the screening of dispersion forces by the solvent and is in evident contrast to the often assumed locality of vdW interactions in solvated systems.
While such a range is unprecedented in the context of solvated systems, similar and larger interaction ranges have already been found for molecular crystals~\cite{Hoja2017} or nanostructures~\cite{Gobre2013}.
From a different point of view, Fig.~\ref{Fig:displacements-FiP35WW}b) represents a radial analysis of the change in the distribution and frequency of electronic fluctuations introduced by embedding the protein in water.
It thus demonstrates that, while the atomistic structure and the \emph{local} dynamics of water typically remain largely unperturbed beyond a few solvation layers~\cite{Bellissent-Funel2016,Persson2018}, the instantaneous electronic structure can indeed indicate the presence of a protein over large distances.
Such long-range collective electronic behavior and the ensuing \emph{non-local} nuclear dynamics could be experimentally measured thanks to recent advances in ultrafast THz spectroscopy~\cite{Ebbinghaus2007,Meister2013,Xu2015}, as we will further discuss below.

%other secondary structures
\subsection*{Effect of secondary structure}
Fip35-WW is a showcase example for the formation of \textbeta-sheets.
To test the general validity of our hypotheses, we carried out the same analysis for the modified villin headpiece, HP-35 NleNle, (formation of \textalpha-helical entities) and the cln025 variant of the \textit{de novo} protein Chignolin (plain \textbeta-hairpin formation).
Our analysis confirmed our early findings for Fip35-WW.
The vdW contribution to the solvation energy reflects the trend of the inverse RMSD from the native structure with a drop of 15\,\% at the hydrophobic collapse.
Again, the protein-water interaction counteracts the destabilizing many-body dispersion effects observed in gas phase and increases the relative stabilization of native states with respect to unfolded structures.
Also, collective plasmon-like electronic fluctuations have been found to show a major contribution to the total and relative solvation energy ($\Delta E_{\mathrm{sol}}^{\mathrm{(low\,\omega)}}$) for both, the hairpin-forming cln025 and the helix-forming HP35-NleNle.

Fig.~\ref{Fig:Summary} summarizes the above-mentioned features and highlights the general validity of the present findings in the biomolecular context.
Independent of the secondary structure to be formed, the vdW solvation energy captures the hydrophobic collapse in the form of a sizable jump in stabilization (``$\Delta E_{\mathrm{sol}}$ at collapse") for all considered proteins.
Also, the consistent increase of the relative stability of native states due to the many-body character of protein-water vdW interactions and the significance of low-frequency, plasmon-like electronic fluctuations (as characterized by their contribution to relative solvation energies, $\Delta E_{\mathrm{sol}}^{\mathrm{(low\,\omega)}}$) turned out to be independent of the final secondary structure.
It is worthwhile to note that the magnitudes of the different quantities are not necessarily representative for secondary structure elements as also the system size varies from 6,000 to 14,000 atoms.
The systematic investigation of the relation between secondary structure elements and the magnitude of the present observations and characteristic features of fluctuation patterns, as shown in Fig.~\ref{Fig:displacements-FiP35WW}a), is beyond the scope of the present publication and subject to ongoing studies.

%CONCLUSION & OUTLOOK
\section*{Discussion and Outlook}
In conclusion, we have shown that many-body dispersion effects lead to a significant relative destabilization ($\approx$ 4.5~kcal/mol for the proteins studied here) of the native state of solvated proteins, when considered in gas phase.
Here, we find that the screening of the instantaneous dipoles due to the surrounding dipole field and many-body interactions contribute in similar parts to the destabilization.
Notably, this effect is of a comparable order of magnitude as estimates for the zero-point vibrational and entropic contribution found for the folding of isolated polypeptides~\cite{Tkatchenko2011,Rossi2013}.
A detailed analysis and comparison of these effects are subject of future work.
The destabilization via many-body dispersion effects can play an important role in explaining why proteins often do not adopt the same folded conformation in the gas phase and in solvation.
It also indicates how the neglect of the inherent many-body character of dispersion interactions in traditional vdW approaches (and molecular mechanics force fields) can lead to a spurious description of \textit{intra}-protein interactions in general.

In aqueous solvation, the vdW contribution to the solute-solvent interaction of (small) proteins captures their hydrophobic collapse and thus represents a viable descriptor for the folding process.
The collapse is accompanied by a jump of about 15\,\% (20--30~kcal/mol) in the vdW solvation energy.
The total electronic energy of solvation, for comparison, does not provide such clear insight (see Supplementary Materials) -- only the free energy of solvation does.
The beyond-pairwise contributions to the protein-water vdW interaction favor folded states and, thus, the many-body aspect of solvation leads to a significant stabilization of native conformations.
This stabilization is the result of a distinct many-body character of the protein-water dispersion interaction in the form of delocalization and a high degree of collectivity of electronic fluctuations across protein and solvent.
This plasmon-like character of vdW interactions in solvated systems can also be pivotal for other quantities, as demonstrated for the protein-water interaction range in Fig.~\ref{Fig:displacements-FiP35WW}b).
Our study shows, that these findings can be generalized for helix-, \textbeta-sheet-, or hairpin-forming proteins and are, thus, independent of secondary structure motifs.
So, an accurate description of solvated proteins in general requires capturing the subtle balance between beyond-pairwise effects on the \textit{intra}-protein vdW interaction (destabilizing native states) and the highly collective, plasmon-like character of protein-water interactions (stabilizing native states).
With increasing system size and complexity, finding this balance without explicit account for the quantum-mechanical many-body nature of vdW interactions is an intricate task and failure to do so can contribute to the fundamental origin of the previously reported~\cite{Best2014,Piana2015} unbalanced description of vdW forces by pairwise molecular mechanics potentials and water models.

In a broader perspective, our findings imply that an effective pairwise-additive treatment of vdW interactions and derivative molecular mechanics potentials can provide accurate energetics for a particular application, but only explicitly quantum-mechanical models, such as the one used here, allow to attain an unbiased microscopic understanding of biomolecular interactions in a more general context.
Based on Fig.~\ref{Fig:displacements-FiP35WW}b), for instance, we expect that for several or larger solutes the approximation of pairwise additivity can fail on a fundamental level.
The persistence and collectivity of electronic fluctuations through the solvent can mediate long-range correlation forces between individual solutes or entities of solvated macromolecules and an effective pairwise description might not be able to reproduce the subtle balance between long-range correlation on all these scales.
As such, plasmon-like interactions can also substantially affect molecular assembly and the formation of tertiary structures.
In addition, complex long-range fluctuations of the electron density are less sensitive to the instantaneous solvent structure and thus thermal fluctuations, which makes them an ideal contender for biomolecular recognition.
In this form of recognition, the solvent provides electron density that serves as a mediator for long-range interaction while the actual atomistic structure and the nuclear dynamics of the solvent do not necessarily have to be altered in the process, which has been concluded from a number of experiments~\cite{Bellissent-Funel2016,Persson2018}.
It is worthwhile to mention, however, that most of these experiments probed rather local interactions and dynamics.
Recent THz-spectroscopy experiments, for instance, show that the presence of a solute can have a considerable effect on the long-timescale dynamics and long-range polarization of water~\cite{Ebbinghaus2007,Meister2013,Xu2015}.
The here observed long-range persistence of electron correlation through aqueous environments will manifest in the slow nuclear dynamics of the system.
Similar behavior has already been observed in crystalline molecular systems, where many-body dispersion effects particularly affect low-frequency (``slow") phonon modes~\cite{Reilly2014,Folpini2017}.
Long-range electronic correlation between (solvated) biomolecules can also form the quantum-mechanical basis for correlated collective nuclear motion within the respective partners.
Such concerted motion is essential for coordinated enzymatic action~\cite{Kurian2016,Kurian2018} or the emergence of coherent molecular vibrations, a promising explanation for long-range recognition through electrodynamic interaction of the resulting oscillating molecular dipoles~\cite{Preto2015,Nardecchia2018}.
Solvent-mediated plasmon-like interactions can also give a quantum-mechanical foundation for the recent proposal by Melkikh and Meijer~\cite{Melkikh2018} of the existence of long-range interactions guiding protein folding, assembly, and organization in cells.
Altogether, our findings in fact apply in a broader context of biomolecular interactions -- not just in the case of protein folding as exemplified here.

Obviously, the stability and functionality of biomolecules is ultimately determined by their free energy.
Hence, this work represents a first step towards a more fundamental understanding of the physics of proteins in water, but to accurately address the implications of plasmon-like features within biological systems, we need to extend our study to free energy at finite temperature.
It is already known that the many-body character of vdW dispersion interactions can significantly soften low-frequency vibrational modes in organic matter, which has a noticeable effect on the system entropy~\cite{Reilly2014,Folpini2017}.
In the case of polymorphs of the aspirin crystal, for instance, many-body dispersion effects introduce a relative entropic stabilization of 0.6~kcal/mol (per molecule in the unit cell) at room temperature~\cite{Reilly2014}.
This can be seen as an estimate for the effect on a protein's relative entropy per residue.
As the dynamics and functionality of a biomolecule can be strongly related to its eigenmodes~\cite{Levy1982,Zheng2017}, the impact on vibrational modes also hints at an unrevealed role of plasmon-like electronic fluctuations for the functionality and coordination in the biochemical machinery.
Besides the proposed roles in long-distance recognition, enzymatic action~\cite{Kurian2016,Kurian2018}, and pharmaceutical activity~\cite{Craddock2017}, this further strengthens the relevance of collective electronic fluctuations in biomolecular systems.
Our DFTB+MBD framework provides a robust formalism for the investigation of such intricate questions as it allows for a fully quantum-mechanical treatment of large-scale systems in atomistic detail.

%Methods
\section*{Methods}\label{subsec:Methods}
For each snapshot along the folding trajectories, we obtain effective atomic polarizabilities, $\alpha$\textsubscript{$A$}, as used within MBD and vdW(TS), in a seamless and non-empirical formalism via net atomic populations as described in Reference~[\citenum{Stoehr2016}].
Electronic structure calculations have been carried out in the Density-Functional Tight-Binding framework with self-consistent charges (SCC-DFTB)~\cite{Porezag1995,Elstner1998} using a locally modified version of DFTB+~\cite{Aradi2007,DFTBvdWGitlab}.
In vdW(TS), the effective isotropic polarizabilities define effective C\textsubscript{6} interaction coefficients for pairwise-additive interaction potentials~\cite{Tkatchenko2009,Stoehr2016}.
In MBD then, $\alpha$\textsubscript{$A$} is subject to short-range electrodynamic screening and then enters the coupled fluctuating dipole model~\cite{Tkatchenko2012}.
So, electron density fluctuations in a system consisting of $N$ atoms are modeled by a set of $N$ three-dimensional, isotropic quantum harmonic oscillators (QHOs) in dipole coupling defined by the Hamiltonian,
\begin{gather}
\mathcal{H}_{\mathrm{MBD}} = \mathcal{T}_{\zeta} + \dfrac{1}{2}\zeta^T\mathcal{V}\,\zeta \qquad ,\label{eq:HMBD}\\
\mathcal{V}^{(i,j)}_{AB} = \omega_A\omega_B\left[\delta_{AB} + \sqrt{\tilde{\alpha}_A\tilde{\alpha}_B}\:\mathcal{D}_{AB}^{(i,j)}\right] \:.\label{eq:VMBD}
\end{gather}
In equation~\eqref{eq:HMBD} $\mathcal{T}$ is the kinetic energy, $\mathcal{V}$ the potential energy matrix, and $\zeta$ the direct sum of mass-weighted displacements of the individual QHOs.
$\mathcal{V}$\textsubscript{$AB$} is defined by the characteristic excitation frequencies, $\omega$, the screened atomic polarizabilities, $\tilde{\alpha}$, and the long-range dipole coupling tensor $\mathcal{D}$\textsubscript{$AB$}~\cite{Tkatchenko2012,Ambrosetti2014}.
In summary, the main approximations within MBD are the coarse-graining of the system's response to an atom-centered framework and the dipole approximation for the coupling between electronic fluctuations.
Unitary transformation of the Hamiltonian~\eqref{eq:HMBD} to a new set of collective variables, $\xi = \mathbf{C}\,\zeta$, such that
\begin{equation}
\mathbf{C}^\dagger\mathcal{V}\,\mathbf{C} = \mathrm{diag}\left\lbrace\tilde{\omega}_i^2\right\rbrace\quad,
\label{eq:CVC}
\end{equation}
transforms equation~\eqref{eq:HMBD} into an uncoupled set of $3N$ one-dimensional QHOs with frequencies $\tilde{\omega}$\textsubscript{$i$} and displacements $\xi$\textsubscript{$i$}.
It therefore provides a model for intrinsic collective charge density fluctuations -- a molecular analogue to the plasmon pseudo-particle in metallic systems.
The dispersion energy has been shown to equal the zero-point interaction energy of this set of QHOs~\cite{Dobson2005,Tkatchenko2013},
given by
\begin{equation}
E_{\mathrm{MBD}} = \frac{1}{2}\sum_{i=1}^{3N} \tilde{\omega}_i - \frac{1}{2}\sum_{A=1}^N\sum_{i=1}^3 \omega_A \quad.
\end{equation}
The contribution of an individual collective electronic fluctuation, $\xi_i$, to the total vdW solvation energy, as defined in equation~\eqref{eq:Esol}, is obtained as the $i$-th element of the vector,
\begin{equation}
\begin{aligned}
\pmb{\varepsilon}_{\mathrm{int}} = \dfrac{1}{2}\mathcal{U}^\dagger\left[\mathbf{C}_{\mathrm{ps}}\right]&\left\lbrace\mathcal{U}\left[\mathbf{C}_{\mathrm{ps}}\right]\pmb{\tilde{\omega}}_{\mathrm{ps}} \right.\\
&\left. - \pmb{\tilde{\omega}}_{\mathrm{sub}} - \mathcal{U}\left[\mathds{1}\right] \left(\pmb{\omega}_{\mathrm{ps}} - \pmb{\omega}_{\mathrm{sub}}\right)\:\right\rbrace\:.
\end{aligned}
\label{eq:eps_int}
\end{equation}
The transformation matrix, $\mathcal{U}[\mathbf{Y}]$, is given by the element-wise absolute square of $(\mathbf{C}$\textsubscript{p}$\,\oplus\,\mathbf{C}$\textsubscript{s}$)$\textsuperscript{$\dagger$}$\,\mathbf{Y}$.
$\mathbf{C}$ corresponds to the transformation matrix used in equation~\eqref{eq:CVC} and $\pmb{\tilde{\omega}}$ to the vector of characteristic frequencies.
The subscript $\mathrm{sub}$ always refers to the direct sum of the corresponding quantity for the isolated protein and the pristine solvent, e.g. $\pmb{\omega}_{\mathrm{sub}} = \pmb{\omega}_{\mathrm{p}}\oplus\pmb{\omega}_{\mathrm{s}}$.
Note that the above transformation preserves energy and thus the sum of all elements in $\pmb{\varepsilon}$\textsubscript{int} equals the total solvation energy.
For further information on the procedure, see Reference~[\citenum{Hermann2017}].
Using the above definitions, we may also obtain the spatial distribution of plasmon-like interactions relevant for the vdW solvation energy.
The radial MBD interaction range, $G_{\mathrm{int}}[\mathrm{MBD}]$, shown in Fig.~\ref{Fig:displacements-FiP35WW}b) is calculated via,
\begin{equation}
G_{\mathrm{int}}[\mathrm{MBD}] = \sum_i \varepsilon_i^{\mathrm{(int)}} \sum_A \delta_{R,R_A} \sum_{j\in A}\Vert\mathbf{C}_{\mathrm{ps}}^{(i,j)}\Vert^2 \label{eq:Gint}
\end{equation}
Here, $\varepsilon_i^{\mathrm{(int)}}$ is the contribution of mode $\xi_i$ to E\textsubscript{sol}, \textit{i.e.} the $i$-th element of $\pmb{\varepsilon}_{\mathrm{int}}$ as defined by equation~\eqref{eq:eps_int}, and $\delta$ corresponds to a generalized Kronecker delta on $\mathds{R}$.
$\mathbf{C}_{\mathrm{ps}}^{(i,j)}$ denotes the elements of the transformation matrix $\mathbf{C}_{\mathrm{ps}}$, which define the x, y, and z components of the subvector of $\xi_i$ that resides on atom $A$.
Note that integrating $G_{\mathrm{int}}[\mathrm{MBD}]$ or $G_{\mathrm{int}}[\mathrm{vdW(TS)}]$ yields the corresponding interaction (solvation) energy.
Further details and in-depth analysis of the physics of plasmon-like electronic fluctuations in solvated (bio)molecular systems will be provided in a future publication.
For additional theoretical background and computational details, see Supplementary Materials.

\section*{Acknowledgements}
The authors thank D.~E.~Shaw~Research for providing the trajectories of cln025 and Fip35-WW used in this work and Jan Hermann for insightful discussions.
M.S. acknowledges financial support from the Fonds National de la Recherche, Luxembourg (AFR PhD Grant CNDTEC).
A.T. was supported by the European Research Council (ERC-CoG Grant BeStMo).
The results presented in this publication have been obtained using the HPC facilities of the University of Luxembourg.

%\section*{Supplementary materials}
%Computational Details\\
%Summary of van der Waals Energetics\\
%Figs. S1 to S8\\
%References \textit{(5--8), (16--18)}

\small
\baselineskip14pt
\setlength{\bibsep}{2pt plus 0.5ex}
\balance

% Bibliography
\bibliographystyle{apsrev}

\end{document}

\maketitle
\tableofcontents
\clearpage

\section{Computational Details and van der Waals Models}
\subsection{Density-Functional Tight-Binding calculations}
Electronic structure calculations have been carried out on the Density-Functional Tight-Binding (DFTB) level of theory using a locally modified MPI-version of DFTB+~[65,66].
For the study of the Fip35 Hpin1 WW-domain and the Nle/Nle mutant of villin HP35 we employed the second-order (self-consistent charges) DFTB method with recent mio-1-1 parameters~[41].
For the calculations on the Chignolin variant cln025 third-order DFTB~[68] (DFTB3) with recent 3ob parameters~[69--71] has been used.
Effective atomic polarizabilities within the DFTB3 framework, as further used in our study, have been tested to comply with the results from the second-order approach. In all calculations we employed a self-consistency criterion of 10\textsuperscript{-5} elementary charges and $\Gamma$-point sampling.

\subsection{van der Waals Dispersion Models}
In our study on many-body dispersion effects on protein energetics and the collectivity of van der Waals (vdW) interactions in solvated biosystems, we have employed a combined approach of DFTB and the Many-Body Dispersion (MBD) formalism~[13,46]. 
The results from MBD have been compared with common pairwise approaches to vdW dispersion interactions, as they are commonly employed in state-of-the-art simulation techniques. Pairwise models are represented by the electronic structure-based vdW(TS)~[42] and Grimme's D2~[43] and D3~[44,45] method.
The central starting point of MBD and vdW(TS) is the definition of effective atomic dipole polarizabilities, $\alpha_{\mathrm{eff}}^{(I)}$, according to the chemical environment. As presented in Reference~[39], these can be derived from DFTB as,
\begin{equation}
\alpha_{\mathrm{eff}}^{(I)} = \frac{\alpha_{\mathrm{free}}^{(I)}}{Z_I}\cdot \sum_{i\in I} \mathbf{P}_{ii} \quad,
\label{eq:alpha_eff}
\end{equation}
where $\mathbf{P}$ is the Mulliken population matrix as obtained from DFTB. In vdW(TS), the obtained effective polarizabilities then define effective C\textsubscript{6} interaction coefficients, which then enter pairwise additive C\textsubscript{6}$\cdot R_{IJ}^{-6}$ potentials. The same functional form is used in Grimme's D2 and D3 schemes. Here, one relies on fixed (D2) or geometry-dependent (D3) interaction coefficients.\\

vdW(TS) calculations have been carried out using a standalone calculator based on '\emph{semp\_disp\_corr.F90}', originally part of CASTEP~[72], within the Atomic Simulation Environment~[73].
Calculations involving D2 or D3 have been performed using the DFTD3 module~[74].
For D3 calculations, Becke-Johnson damping as proposed in Reference~[45] has been used.\\
 
In MBD, $\alpha_{\mathrm{eff}}^{(I)}$ is first subject to self-consistent, electrodynamic screening (SCS) to account for the presence of the surrounding fluctuating atomic dipoles and obtain effective, screened atomic polarizabilities $\tilde{\alpha}_{\mathrm{eff}}^{(I)}$. This is achieved by inverting the Dyson-like equation for the dynamic polarizabilities,
\begin{equation}
\tilde{\alpha}_{\mathrm{eff}}^{(I)}\left(i\omega\right) = \sum_{J} \mathbf{B}_{IJ} \quad,\qquad\mathrm{with}\quad\mathbf{B} = \left[\mathbf{A}^{-1}(i\omega) + \mathbf{T}_{\mathrm{gg}}\right]^{-1} \quad,
\end{equation}
where $\mathbf{A}(i\omega)$ is the diagonal matrix of the dynamic polarizabilities $\alpha_{\mathrm{eff}}^{(I)}(i\omega)$ at imaginary frequency $i\omega$ and $\mathbf{T}_{\mathrm{gg}}$ is the effective short-range dipole potential constructed from the Coulomb interaction of two overlapping Gaussian charge densities. In the MBD formalism we model the interaction of intrinsic electronic fluctuations in form of a set of dipole-coupled Quantum Harmonic Oscillators (QHOs). In accordance with this, the dynamic polarizability and effective excitation frequency, $\omega_I$, are defined as
\begin{equation}
\alpha_{\mathrm{eff}}^{(I)}(i\omega) = \frac{\alpha_{\mathrm{eff}}^{(I)}}{1 + \left(\frac{\omega}{\omega_I}\right)^2} \qquad\mathrm{and}\qquad \omega_I = \frac{4}{3}\frac{\alpha_{\mathrm{eff}}^{(I)}}{\alpha_{\mathrm{free}}^{(I)}}\frac{C_{6,\mathrm{free}}^{(II)}}{\alpha_{\mathrm{free}}^{(I)}} \quad,
\end{equation}
respectively. The effective, screened polarizabilities then define the MBD Hamiltonian,
\begin{equation}
\mathcal{H}_{\mathrm{MBD}} = \mathcal{T}_{\zeta} + \dfrac{1}{2}\zeta^T\mathcal{V}\,\zeta \quad,\qquad \mathrm{with}\quad \mathcal{V}^{(k,l)}_{IJ} = \omega_I\omega_J\left[\delta_{IJ} + \left(1 - \delta_{IJ}\right)\sqrt{\tilde{\alpha}_I\tilde{\alpha}_J}\:\mathcal{D}_{IJ}^{(k,l)}\right] \:.
\label{eq:HMBD}
\end{equation}
In eq.~[\ref{eq:HMBD}], $\mathcal{T}$ is the kinetic energy, $\mathcal{V}$ the potential energy matrix, and $\zeta$ the direct sum of mass-weighted displacements of the individual QHOs. $\mathcal{V}$\textsubscript{$IJ$} is defined by the characteristic excitation frequencies, screened polarizabilities, and a damped dipole coupling tensor (Fermi-damping) $\mathcal{D}$\textsubscript{$IJ$}~[13,46]. 
All MBD and vdW(TS){@}SCS calculations have been performed using a self-written implementation.

\clearpage

\section{Summary of Gas-phase Energetics for Fip35 Hpin1 WW-domain}
\begin{figure}[!h]
\centering
\includegraphics[scale=0.61]{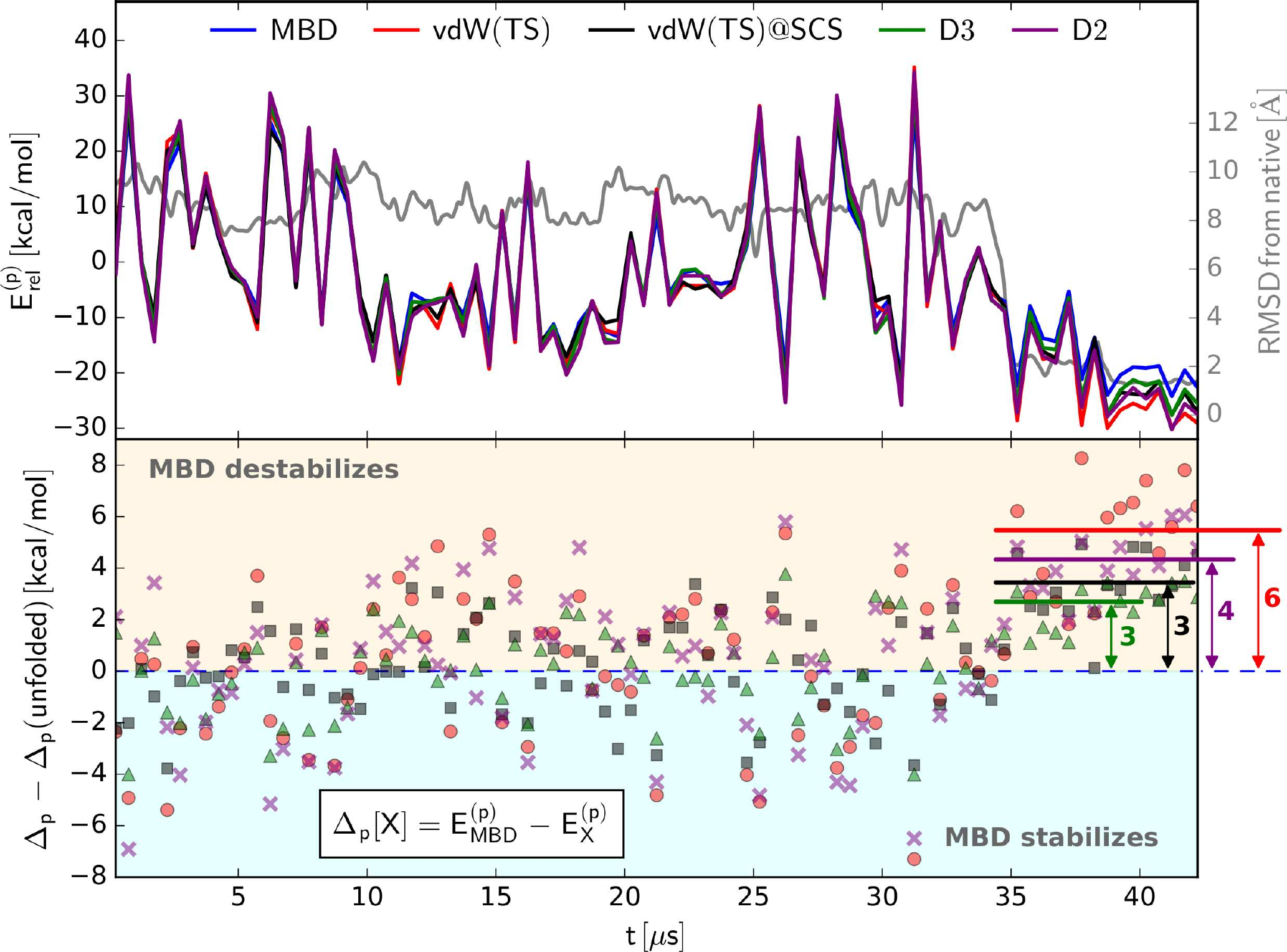}
\caption{van der Waals dispersion energy of Fip35 Hpin1 WW-domain in gas-phase as obtained with all considered van der Waals models (top). Beyond pairwise effects on gas-phase energetics (bottom).}
\label{Fig:Ep-Fip35}
\end{figure}

As can be seen from Fig.~\ref{Fig:Ep-Fip35}, all considered vdW models show comparable behavior along the folding trajectory. The main differences between the approaches can be found for the relative stability of native states with respect to unfolded conformations. Hereby, the pairwise models overestimate the stability as seen from Fig.~\ref{Fig:Ep-Fip35}(bottom). The average over-stabilization is 3\,kcal/mol in Grimme's D3 and vdW(TS) with screened C\textsubscript{6} interaction coefficients, ``vdW(TS)@SCS''. For D2 and conventional vdW(TS) we find an overestimation of the relative stability by 4 and 6\,kcal/mol, respectively.
\clearpage

\section{van der Waals Energetics in Detail}
\subsection{Fip35 Hpin1 WW-domain}
\begin{figure}[!h]
\centering
\includegraphics[scale=0.8]{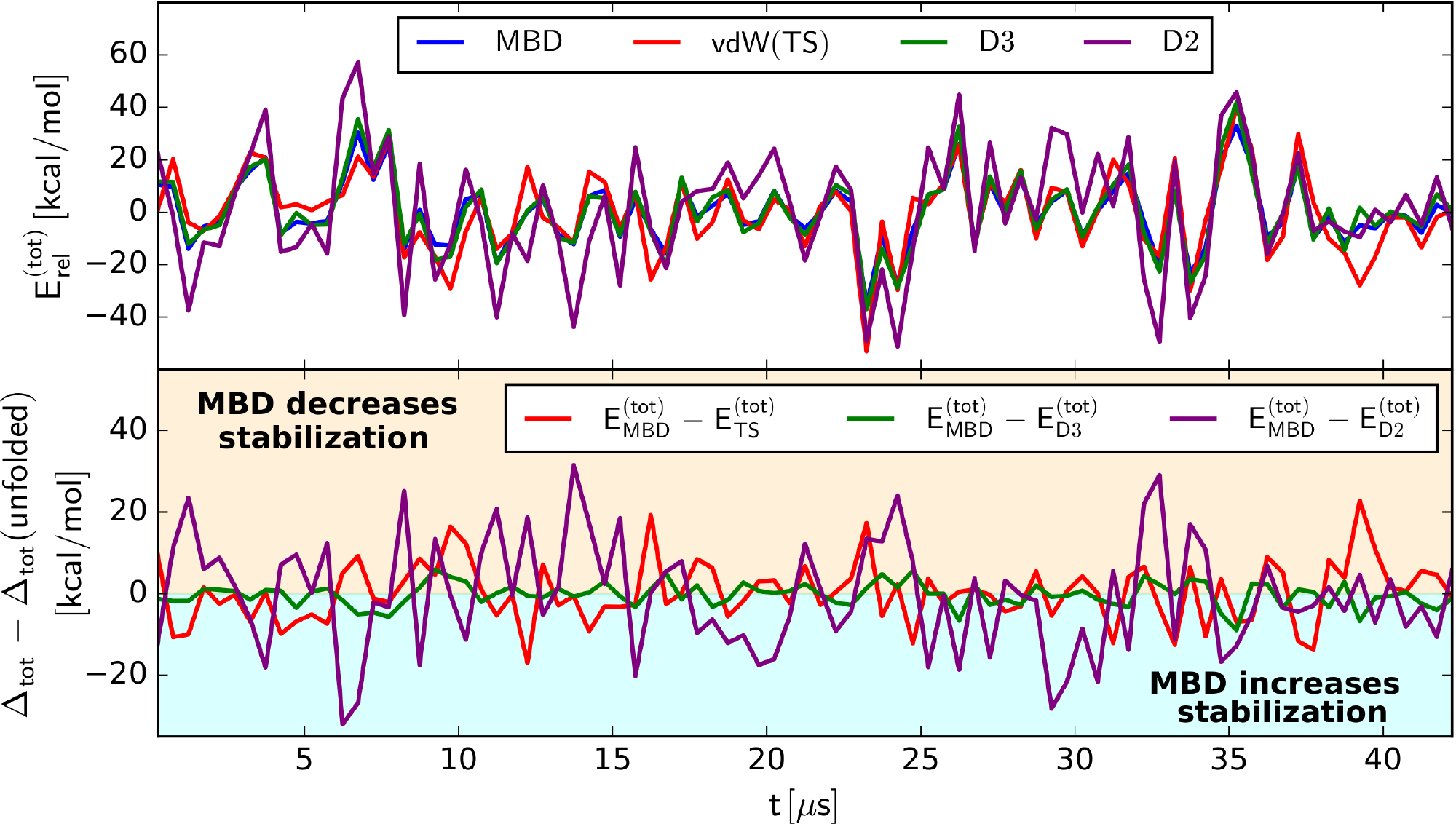}
\caption{Total relative van der Waals dispersion energy of Fip35 Hpin1 WW-domain in solvation as obtained using the many-body dispersion formalism and the pairwise approaches vdW(TS), D2, and D3 (top). Beyond pairwise effects on relative van der Waals energetics of the Fip35 Hpin1 WW-domain in aqueous solvation (bottom).}
\label{Fig:Etot-Fip35}
\end{figure}

Upon embedding in an aqueous environment, we do not observe systematic differences in the relative vdW energetics (\emph{cf.} Fig.~\ref{Fig:Etot-Fip35}). Also in the case of the pristine solvent, the pairwise models do not show consistent deviations from MBD.
Interestingly, for the geometry-based D2 and D3, we additionally find a subtle relative over-stabilization of the pristine solvent corresponding to native state conformations. These add up with minor, yet non-systematic, discrepancies in the total vdW energy and ultimately lead to a overall underestimation of the relative solvation energy of native conformations by 12\,kcal/mol for D2 and 7\,kcal/mol for D3 (see main manuscript). The latter represents a surprising shortcoming when compared to the electronic structure-based vdW(TS) as D3 outperforms all other pairwise models for gas-phase vdW energies (\emph{cf.} Fig.~\ref{Fig:Ep-Fip35}). The spurious description of the pure solvent in the geometry-motivated models can most likely be ascribed to an inaccurate description of vdW interactions involving hydrogen atoms located near the cavity and thus mainly affects the pristine solvent. We conclude that, for an accurate description of edge effects in water, it is essential to include electronic-structure effects.
In terms of the description of protein-water vdW interactions, ultimately, the pairwise approaches systematically lack a relative stabilization of 5 to 12\,kcal/mol due to neglect of beyond-pairwise interaction terms.
\clearpage

\subsection{Chignolin variant ``cln025''}
\begin{figure}[!h]
\centering
\includegraphics[scale=1.]{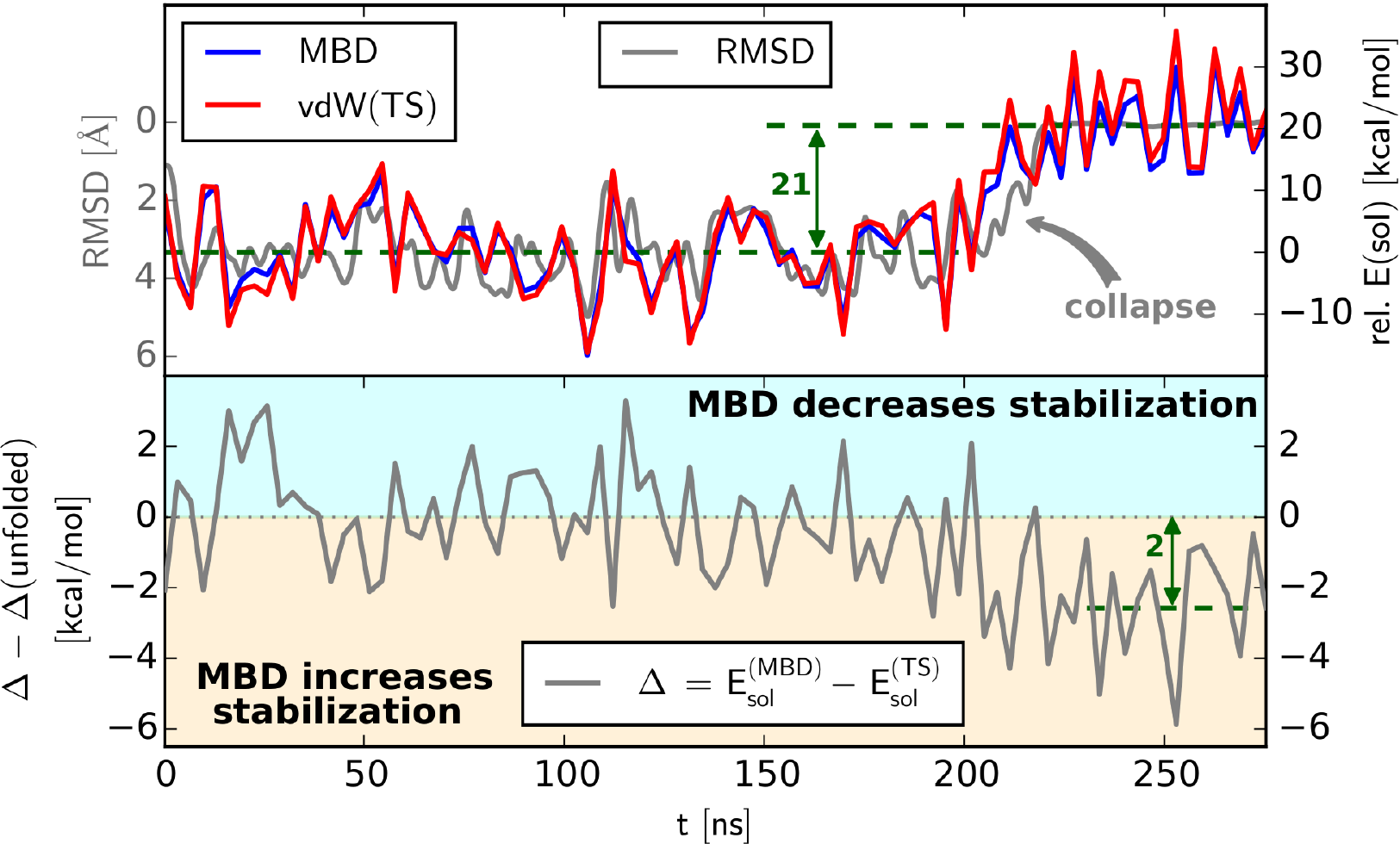}
\caption{Relative van der Waals solvation energy of cln025 as obtained with the many-body dispersion model and the pairwise vdW(TS) approach (top). Beyond pairwise contributions to van der Waals solvation energy as given by difference between many-body and pairwise treatment (bottom).}
\label{Fig:Esol-cln025}
\end{figure}

\begin{figure}[!h]
\centering
\includegraphics[scale=1.]{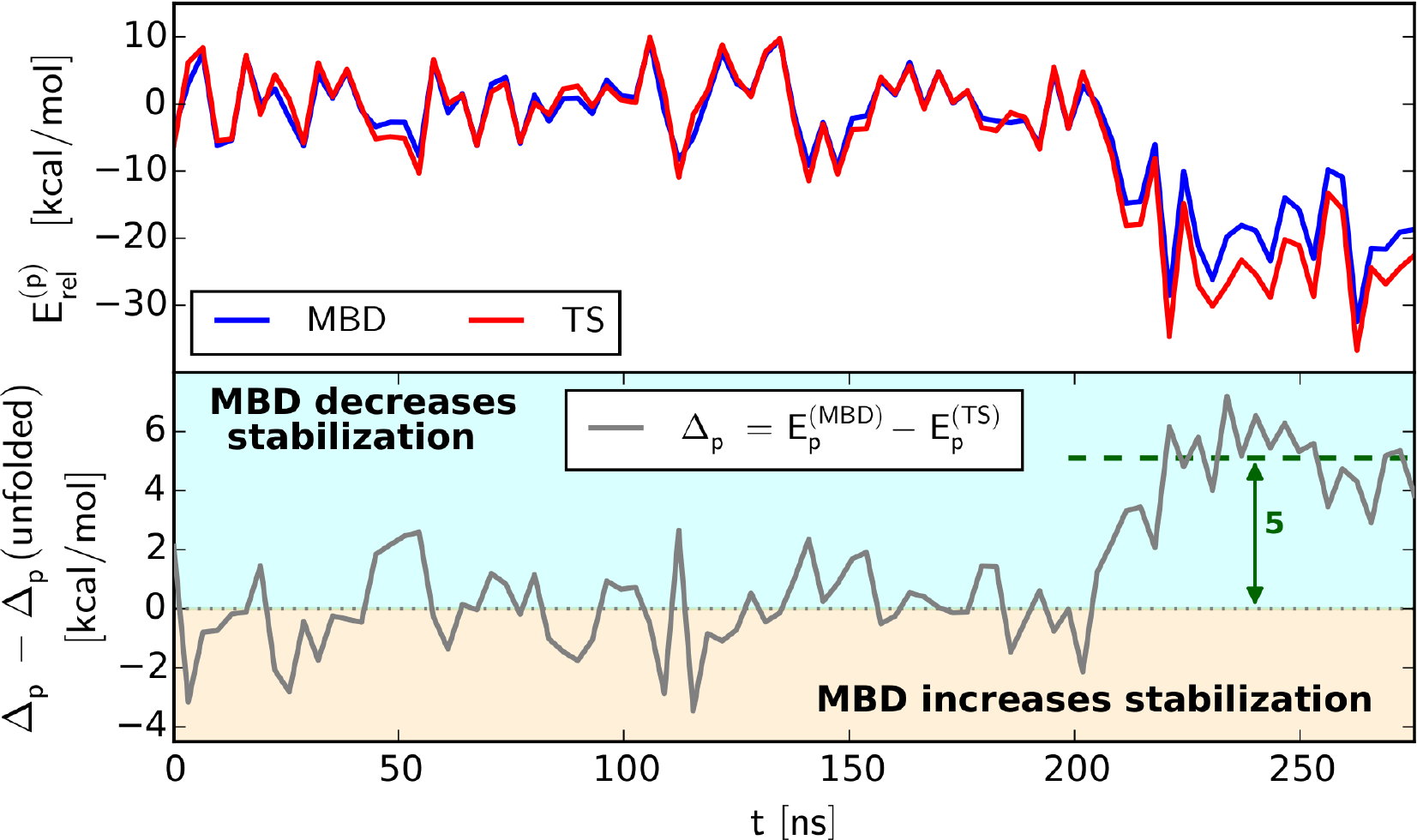}
\caption{Relative van der Waals energy of the Chignolin variant cln025 in absence of solvent as obtained with the many-body (MBD) and pairwise (vdW(TS)) treatment of dispersion interaction (top). Beyond pairwise contributions to relative van der Waals energetics (bottom).}
\label{Fig:Ep-cln025}
\end{figure}

For the ``cln025'' variant of the \emph{de novo} protein Chignolin, we observe a similar behavior for the gas-phase energetics. The dispersion interaction energy is maximized when going to the native state and the pairwise vdW model vdW(TS) overestimates the relative stability of the native state of cln025 in the absence of water in comparison to MBD by 5\,kcal/mol (see Fig.~\ref{Fig:Ep-cln025}). Upon embedding in water, we still find a slight destabilization of native states via many-body dispersion effects (2\,kcal/mol), while for the pure solvent we did not observe a statistically relevant change in the relative vdW energy along the folding trajectory. The discrepancy in the relative vdW solvation energy of cln025 between pairwise and many-body treatment, as depicted in Fig~\ref{Fig:Esol-cln025}, is thus governed by the neglect of many-body dispersion effects on \textit{intra}-protein vdW energetics during folding and a slight destabilization of native states of cln025 in solvation via many-body effects.
The increase of the relative stabilization through beyond-pairwise protein-water vdW interactions amounts to 2\,kcal/mol.
\clearpage

\subsection{Villin headpiece 35 Nle/Nle mutant}
For HP35-NleNle, many-body dispersion effects destabilize the native state in absence of water, \textit{i.e.} reduce the \textit{intra}-protein vdW interaction, by 4\,kcal/mol in comparison to the pairwise vdW(TS). 
For solvated HP35-NleNle and the pure solvent we did not observe a considerable consistent many-body effect on the energetics.
In sum, the pairwise formalism underestimates the protein-water vdW interaction of HP35-NleNle by about 3\,kcal/mol when compared to a full many-body treatment.
\clearpage

\section{Correlation and Rescaling of van der Waals Solvation Energies}
\begin{figure}[!h]
\centering
\includegraphics[scale=.9]{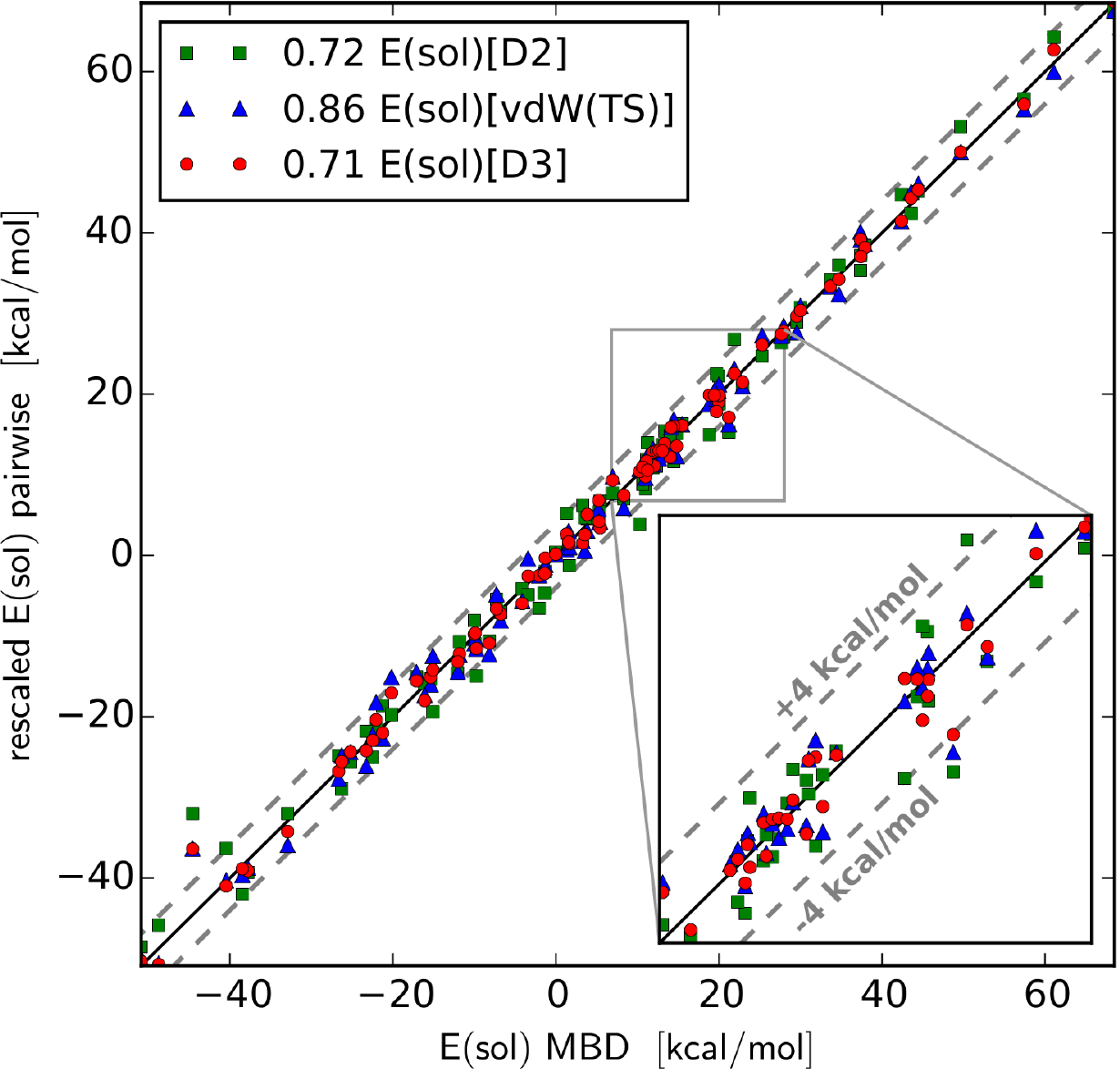}
\caption{Correlation of rescaled relative van der Waals solvation energies as obtained from pairwise models in comparison to the results obtained from many-body treatment.}
\label{Fig:Esol-correlation}
\end{figure}

As can be seen from Fig.~\ref{Fig:Ep-Fip35} and the main manuscript, the overestimation of vdW energies increases with the absolute vdW interaction energy. Correspondingly, a simple rescaling of the pairwise approaches considerably improves the agreement with the many-body treatment. Fig.~\ref{Fig:Esol-correlation} shows the correlation between such optimally rescaled vdW solvation energies and MBD. The obtained rescaling factors show that relying on electronic-structure based C\textsubscript{6} interaction coefficients as done within vdW(TS) provides the best estimate for vdW solvation energies. This can mainly be attributed to the description of the pure solvent as the geometry-based D2 and D3 methods outperform vdW(TS) for gas-phase energetics (\emph{vide supra}). Despite the overall improvement, the deviation between the optimally rescaled pairwise approaches and MBD still regularly exceeds 4\,kcal/mol. Furthermore, the optimal rescaling factors are highly system- and method-dependent and can only be obtained as an \emph{a posteriori} correction.

\section{Total Electronic Energy of Solvation}
\begin{figure}[!h]
\centering
\includegraphics[scale=.9]{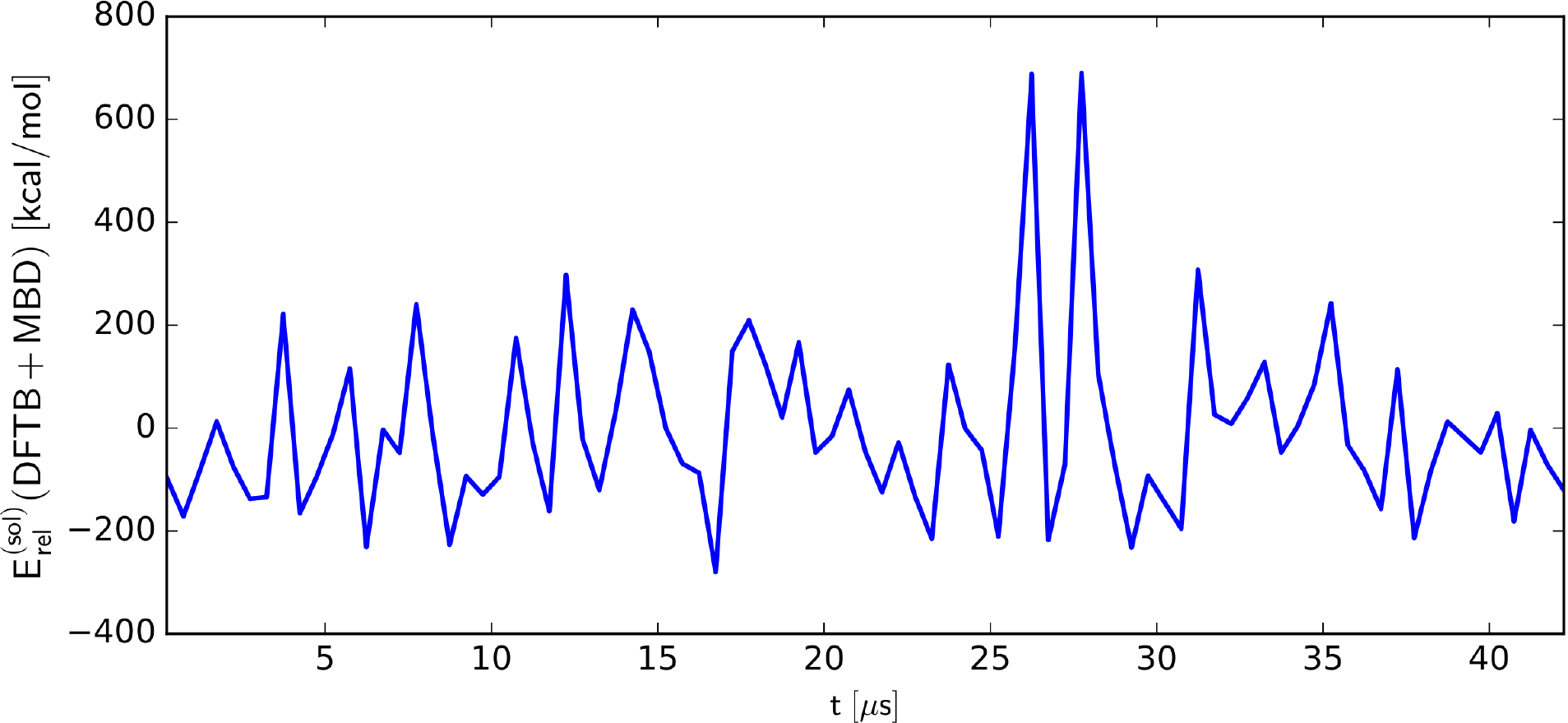}
\caption{Total electronic energy of solvation of the Fip35 Hpin1 WW-domain as obtained with density-functional tight-binding in conjunction with the many-body dispersion model.}
\label{Fig:Esoltot-Fip35}
\end{figure}

\begin{figure}[!h]
\centering
\includegraphics[scale=.86]{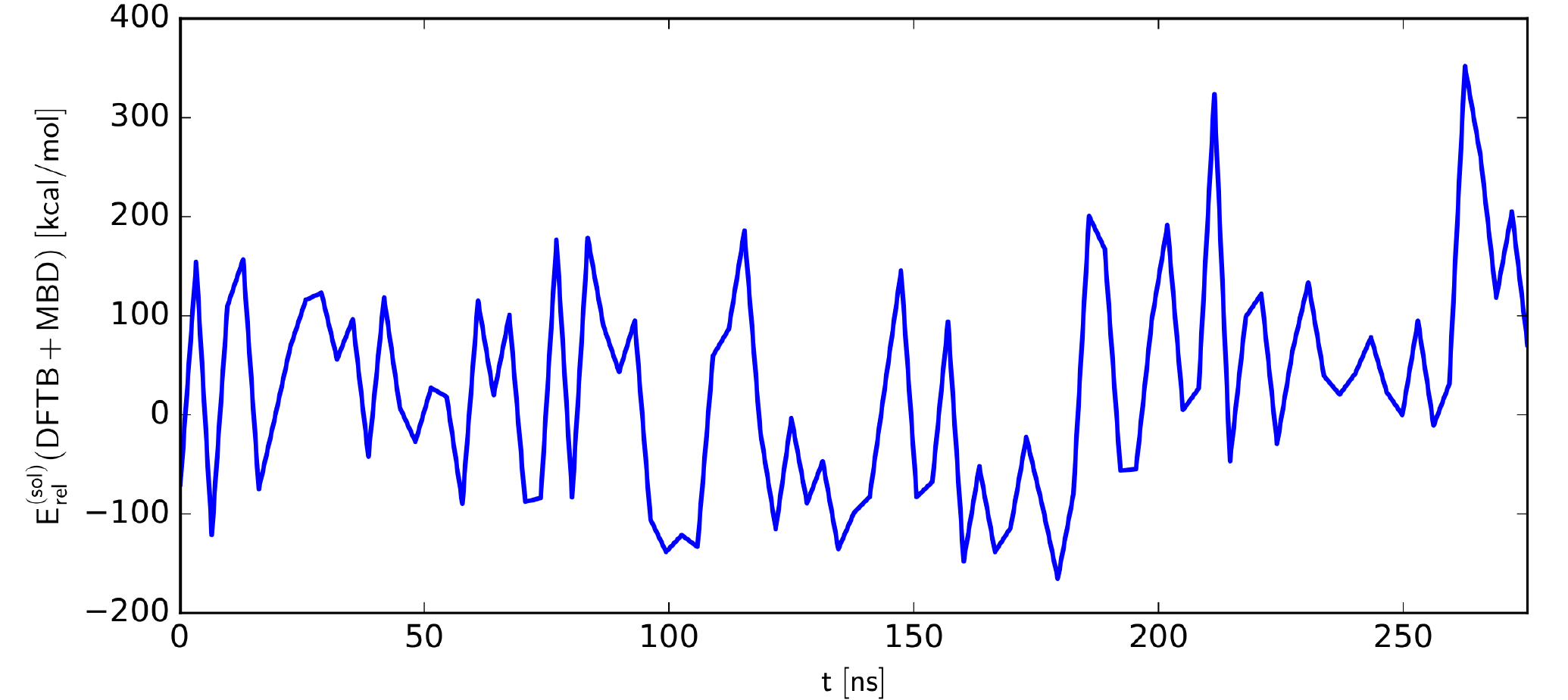}
\caption{Total electronic energy of solvation of the Chignolin variant ``cln025'' as obtained with density-functional tight-binding in conjunction with the many-body dispersion model.}
\label{Fig:Esoltot-cln025}
\end{figure}

In contrast to the vdW solvation energy, the total electronic energy of solvation does not provide a clear-cut distinction between folded and unfolded states (\emph{cf.} Fig.~\ref{Fig:Esoltot-Fip35} and \ref{Fig:Esoltot-cln025}). This is not surprising as the, in the end decisive, free energy of a solvated molecule has a large entropic component and it is known that the hydrophobic effect mainly arises from entropic contributions~[5].
In the case of cln025, we do observe a slight shift in the total electronic energy of solvation when comparing folded and unfolded states, see Fig.~\ref{Fig:Esoltot-cln025}. Considering the absolute spread of the solvation energy, however, this shift is less clear as for its mere vdW component (Fig.~\ref{Fig:Esol-cln025}) and coincides with the spread of total solvation energies of the unfolded conformations.
\clearpage